\documentclass[12pt]{article}

\usepackage{scicite}

\usepackage[colorlinks=true,linkcolor=blue,citecolor=blue,urlcolor=blue]{hyperref}

\usepackage{graphicx}
\usepackage{tikz} %to draw graphics
\usepackage{amsmath}
\usepackage{color}
\usepackage{amsmath}
\usepackage{amssymb}
\usepackage{verbatim}
\usepackage{latexsym}
\usepackage{enumerate} % to change the numbering of 'enumerate'
\usepackage{bm} % for bold face mathematical symbols

\usepackage[caption=false]{subfig}
\usepackage{multirow}
\usepackage{booktabs}

\usepackage{times}
\newenvironment{sciabstract}{%
\begin{quote} \bf}
{\end{quote}}

\title{Topological phonons in oxide perovskites controlled by light}

\author
{Bo Peng,$^{1}$ Yuchen Hu,$^{1,2}$
Shuichi Murakami$^{3,4}$, \\
Tiantian Zhang,$^{3,4\ast}$ Bartomeu Monserrat$^{1,5\dag}$\\
\\
\normalsize{$^{1}$Cavendish Laboratory, University of Cambridge,}\\
\normalsize{J.\,J.\,Thomson Avenue, Cambridge CB3 0HE, United Kingdom}\\
\normalsize{$^{2}$Department of Chemistry, University of Cambridge,}\\
\normalsize{Lensfield Road,, Cambridge CB2 1EW, United Kingdom}\\
\normalsize{$^{3}$Department of Physics, Tokyo Institute of Technology,}\\
\normalsize{Ookayama, Meguro-ku, Tokyo 152-8551, Japan}\\
\normalsize{$^{4}$Tokodai Institute for Element Strategy, Tokyo Institute of Technology,}\\
\normalsize{ Nagatsuta, Midori-ku, Yokohama, Kanagawa 226-8503, Japan}\\
\normalsize{$^{5}$Department of Materials Science and Metallurgy, University of Cambridge,}\\
\normalsize{27 Charles Babbage Road, Cambridge CB3 0FS, United Kingdom}\\
\\
\normalsize{$^{\ast\dag}$To whom correspondence should be addressed;}\\
\normalsize{E-mail: $^{\ast}$ ttzhang@stat.phys.titech.ac.jp; $^{\dag}$ bm418@cam.ac.uk}}

\date{}

\begin{document} 

% Double-space the manuscript.

\baselineskip24pt

% Make the title.

\maketitle

% Place your abstract within the special {sciabstract} environment.

\begin{sciabstract}

%Perovskite oxides exhibit a rich variety of structural phases with different functionalities that find multiple technological applications. We find a series of \textcolor{red}{noncentrosymmetric oxide perovskites} that exhibit three types of topological states in their phonon spectra: nodal rings, nodal lines, and Weyl points. Remarkably, in the tetragonal phase of BaTiO$_3$, PbTiO$_3$, and SrTiO$_3$, all these topological phonons can simultaneously emerge when stabilized by photoexcitation, whereas the tetragonal phase stabilized by thermal fluctuations only hosts a more limited set of topological phonon states. Additionally, we find that the photoexcited carrier concentration can be used to tune the topological phonon states and induce topological transitions even without associated structural {\color{red} phase} changes. \textcolor{red}{More generally, we also find topological phonons in multiple structural phases of the perovskites, \textit{e.g.}, orthorhombic and rhombohedral phases of BaTiO$_3$, and also with other tuning parameters such as strain and temperature.} Our proposal, for the first time, of a photo-induced tunable topological phase transition in a phononic system may provide new avenues for controllable transient topological states in light control applications.

Perovskite oxides exhibit a rich variety of structural phases hosting different physical phenomena that generate multiple technological applications. We find that topological phonons -- nodal rings, nodal lines, and Weyl points -- are ubiquitous in oxide perovskites in terms of structures (tetragonal, orthorhombic, and rhombohedral), compounds (BaTiO$_3$, PbTiO$_3$, and SrTiO$_3$), and external conditions (photoexcitation, strain, and temperature). In particular, in the tetragonal phase of these compounds all types of topological phonons can simultaneously emerge when stabilized by photoexcitation, whereas the tetragonal phase stabilized by thermal fluctuations only hosts a more limited set of topological phonon states. Additionally, we find that the photoexcited carrier concentration can be used to tune the topological phonon states and induce topological transitions even without associated structural phase changes. Overall, we propose oxide perovskites as a versatile platform in which to study topological phonons and their manipulation with light.

% in light-control related fields.
% new avenues for light-controlled neuromorphic computing in phononic systems.

\end{sciabstract}

\section*{Introduction}

{In the family of topological materials, semimetals have an important place in the study of topological order because they exhibit topologically protected surface states \cite{Heikkila2011}, an anomalous bulk transport phenomenon known as the ``quantum anomaly'' \cite{Son2013,Xiong2015}, diversified classifications like nodal rings and lines \cite{Weng2015b,Bernevig2018,Fang2015} and Weyl and Dirac points \cite{Wang2012a,Armitage2018,Xu2015,Weng2015,Yang2015}, and they serve as a platform for obtaining various other topological states, such as topological (crystalline) insulators \cite{Yang2018} and Chern insulators \cite{He2017}. Beyond electronic structures, topological degeneracies in the spectra of other quasiparticles such as excitons, magnons, and phonons, have drawn wide attention in the past decade \cite{Wu2017a,Li2016d,Stenull2016,Liu2017a,He2018}. One advantage of the latter group is that excitation of these quasiparticles depends only on the energy of the external probe, rather than being restricted to the bands near the Fermi level like in the case of electrons. Among these quasiparticles, phonons are of particular interest as basic emergent bosonic excitations associated with lattice vibrations \cite{Zhang2018a,Miao2018,Li2018a,Xia2019,Zhang2019c,Liu2020}, and contribute to many physical processes, such as conventional superconductivity \cite{Drozdov2015,Ahadi2019} and the thermal Hall effect \cite{Murakami2016,Chen2020,Hamada2020,Li2020}. Thus, studies of topological phonons, and especially on the control of topological phonons, are becoming an important field in condensed matter physics.}

{Oxide perovskites have gained popularity as a material system in the past decades due to hosting multiple competing phases including ferroelectric \cite{Cohen1992}, magnetic \cite{Moritomo1997}, or superconducting \cite{Ahadi2019}, and also exotic excitations such as skyrmions \cite{Nahas2015,Das2019}. 
Additionally, these competing phases are particularly sensitive to external stimuli, including temperature \cite{Mahmoud2014}, strain\cite {Ahn2004}, pressure \cite{Shirako2009}, and composition \cite{Rusevich2019}. More recently, theoretical and experimental evidence has demonstrated that light is yet another means by which it is possible to control the crystal symmetry of perovskites \cite{Paillard2019,Nova2019,Ahn2019,Porer2019,Juraschek2017}. As topological order is intimately related to symmetry, the versatility of the perovskite family makes these structures promising platforms to also explore topological properties. %{\color{red} Topological orders such as skyrmions in oxide perovskites have been the main topic of research in the ferroics community \cite{Nahas2015,Das2019}. 
%{Given the versatility of perovskite oxides and their varied uses in technological applications, finding topological order in these materials would greatly expand their applicability as a material platform.} 
However, most oxide perovskites are large gap insulators (for example BaTiO$_{3}$, SrTiO$_{3}$, and PbTiO$_{3}$), and as a consequence they exhibit no electronic topological states.}

%Among this structural variety, the noncentrosymmetric tetragonal phase has received significant attention due to the intriguing physical properties it exhibits, such as ferroelectricity \cite{KAY1947,Matthias1948} and piezoelectricity \cite{Zheng2012}, and the variety of ways with which the phase can be stabilized, including temperature, pressure, and strain \cite{Mahmoud2014,Cai2018,Ahn2004,Xiong1986,Moritomo1997,Shirako2009,Rusevich2019}. 

%A until recently has theoretical and experimental evidence shown that light can control the crystal symmetry of perovskites by stabilizing the soft phonon mode \cite{Paillard2019,Nova2019,Ahn2019,Tsai2018,Porer2019}.

In this work, we show that the phonon spectrum of multiple noncentrosymmetric perovskites can host three types of topological states: topological nodal rings, nodal lines, and Weyl points; suggesting that topological phonons are pervasive in different structural phases of oxide perovskites. Using the tetragonal BaTiO$_3$ phase as a prototype, we show that all these types of topological phonon emerge simultaneously when this phase is stabilized by photoexcitation. Additionally, topological order can be controlled with the photoexcited carrier density, driving topological transitions without any associated structural phase change, including the creation and annihilation of Weyl phonons and switching between nodal-ring and nodal-line phonons. In contrast, when the tetragonal phase of BaTiO$_3$ is stabilized by thermal fluctuations, it only exhibits a more limited number of topological states. This is because the long-range Coulomb interaction leads to a large energy splitting between the longitudinal and transverse optical phonons (LO-TO splitting), reducing the possibility of topological degeneracies between these bands. The photoexcitation route eliminates this problem because photoexcited carriers can screen long-range interactions.

\section*{Results}

\subsection*{Crystal structures and structural phase transitions in BaTiO$_3$}

\begin{figure*}
\centering
\includegraphics[width=\linewidth]{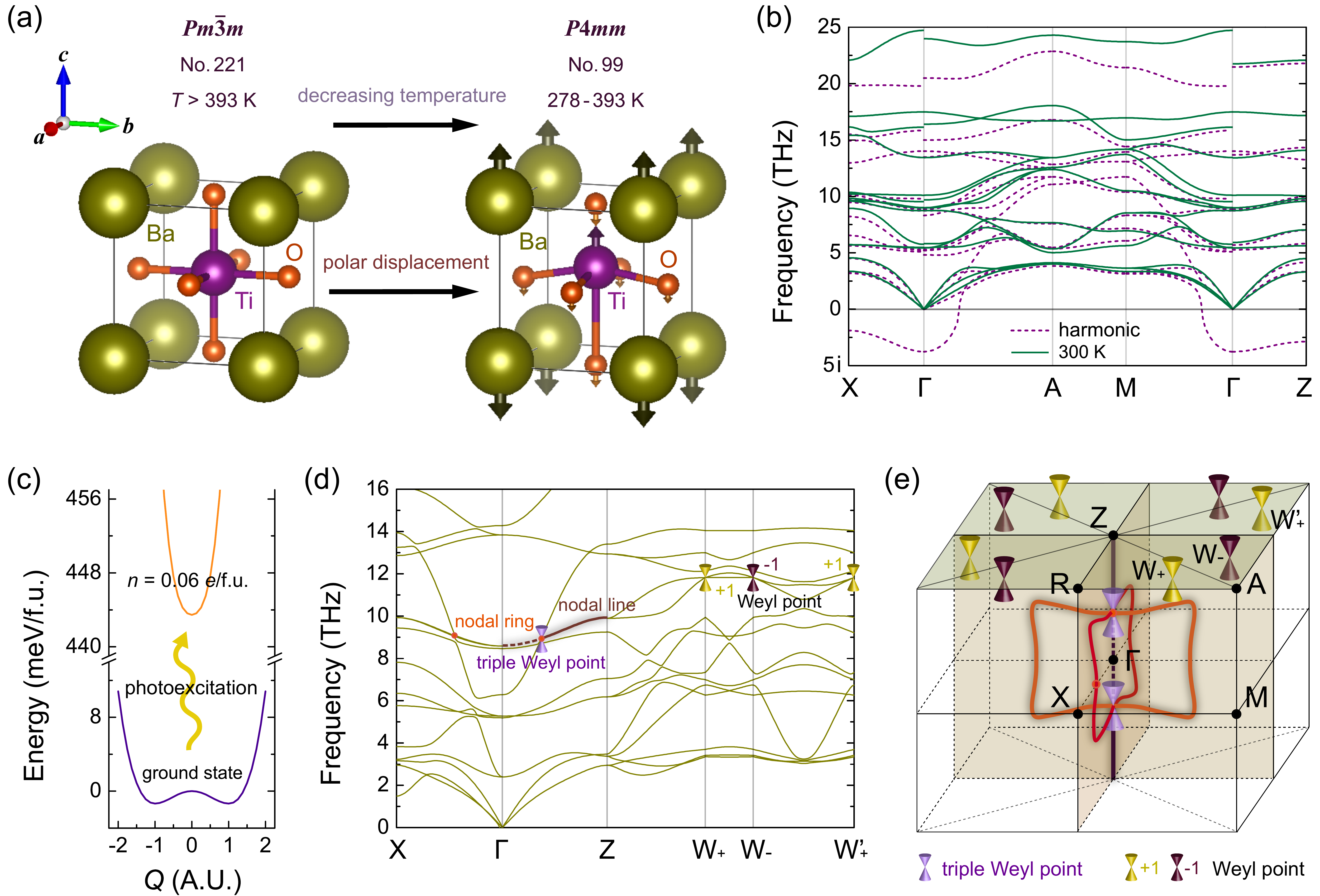}
\caption{\textbf{Crystal structure and phonon dispersion of tetragonal BaTiO$_3$.} (a) Crystal structures of the high-temperature cubic phase (No.~221, $Pm\bar{3}m$) and of the room-temperature tetragonal phase (No.~99, $P4mm$) of BaTiO$_3$. (b) Phonon dispersion of tetragonal BaTiO$_3$ at the harmonic level and at $300$\,K including anharmonic vibrations. (c) Ground-state and excited-state potential energy surfaces along the $\Gamma$ point phonon mode that is imaginary within the harmonic approximation. (d) Phonon dispersion of tetragonal BaTiO$_3$ at $n=0.06$ $e$/f.u. (e) Bulk Brillouin zone at $n=0.06$ $e$/f.u. with two nodal rings on the $q_{x}=0$ and $q_{y}=0$ mirror planes (orange circles, formed by the 10$^{th}$ and 11$^{th}$ bands), one nodal line along the $\Gamma$-Z high-symmetry line (brown line, formed by the 10$^{th}$ and 11$^{th}$ bands), one pair of triple Weyl points at the 
intersection points between the nodal rings and nodal line (violet cones, formed by the 10$^{th}$, 11$^{th}$ and 12$^{th}$ bands), and four pairs of Weyl points on the $q_z$ = $\pi$ plane (purple and yellow cones, formed by the 10$^{th}$ and 11$^{th}$ bands).
}
\label{f1} 
\end{figure*}

As one of the most studied perovskite materials, BaTiO$_3$ has a cubic ABO$_3$-type crystal structure at temperatures above $393$\,K, with the A cation (Ba$^{2+}$) at the corners and the B cation (Ti$^{4+}$) at the center of an octahedral cage of oxygen atoms. Upon cooling, BaTiO$_3$ undergoes an inversion symmetry-breaking phase transition, giving rise to a tetragonal phase with a deformation along the [001] direction \cite{KAY1947}. As shown in Fig.~\ref{f1}(a), the phase transition breaks inversion symmetry by the polar displacements of Ti and O atoms along the $z$ direction, which changes the space group from $Pm\bar{3}m$ (No.~221) to $P4mm$ (No.~99). With further cooling, an orthorhombic $Amm2$ phase appears below $278$\,K and a rhombohedral $R3m$ phase follows below $183$\,K. As a result of these temperature-driven phase transitions, the room temperature tetragonal phase exhibits imaginary harmonic phonon modes, as shown in Fig.~\ref{f1}(b). The inclusion of anharmonic vibrations \cite{Souvatzis2008} stabilizes the tetragonal structure at $300$\,K, also shown in Fig.~\ref{f1}(b).

In addition to temperature, multiple strategies have traditionally been proposed to control structural phase transitions in perovskites, including strain\cite {Ahn2004}, pressure \cite{Moritomo1997,Shirako2009}, and composition \cite{Rusevich2019}. More recently, it has been shown that photoexcitation provides an alternative route to stabilizing multiple perovskite phases \cite{Paillard2019,Nova2019,Ahn2019,Porer2019,Juraschek2017}. Photoexcitation is relatively economical, and can be easier to control compared to other strategies such as heating to change temperature, material synthesis to change composition or strain, or external pressure. Inspired by these recent discoveries, we find that photoexcitation can also stabilize the imaginary phonon modes of the tetragonal phase of BaTiO$_3$: Fig.~\ref{f1}(c) shows that the double well potential energy curve along the imaginary phonon mode of amplitude $Q$ at the $\Gamma$ point becomes a single well upon illumination, indicating the stabilization of the crystal structure. The underlying physics is that changing the carrier concentration $n$ induced by photoexcitation leads to changes in the potential energy experienced by the ions, which in turn changes the interatomic force constants and the phonon dispersion. For $0.04<n<0.10$ $e$/f.u., the $P4mm$ phase is not only dynamically stable, but it is also thermodynamically more stable than the cubic $Pm\bar{3}m$ phase; whereas for $n>0.10$ $e$/f.u., the $P4mm$ phase relaxes to the cubic phase as the latter becomes thermodynamically more stable \cite{Paillard2019}. The relative energy between the two phases and their corresponding lattice constants under illumination are shown in the Supplementary Materials. Figure~\ref{f1}(d) shows the phonon dispersion with a photoexcited carrier density of $n=0.06$ $e$/f.u., confirming the dynamical stabilization of the tetragonal phase.

%The two phonon dispersions in Figure~\ref{f1}(b) and (d) indicate that the tetragonal phase can be stabilized by photoexcitation or alternatively at temperatures between $278$-$393$\,K. However, the difference between the two phonon dispersions is that without photoexcitation, the long-range Coulomb interaction leads to a large LO-TO splitting \cite{Zhong1994a}, which tends to lift the band degeneracy, as shown in Fig.~\ref{f1}(b). Consequently, no Weyl phonons can be observed in room-temperature tetragonal BaTiO$_3$ (although the presence of nodal-line and nodal-ring phonons is confirmed in the Supplementary materials). In contrast, under photoexcitation, the photoexcited carriers result in strong free-carrier screening and the static dielectric constant becomes infinitely large even with extremely low $n$. As a result, the long-range Coulomb interactions are fully screened and the LO-TO splitting in the excited-state phonon dispersion can be ignored. Therefore, photoexcitation is the only way to obtain Weyl phonons in tetragonal BaTiO$_3$ (discuessed later).
%

\subsection*{Topological phonons in BaTiO$_3$}

\begin{figure*}[h]
\centering
\includegraphics[width=\linewidth]{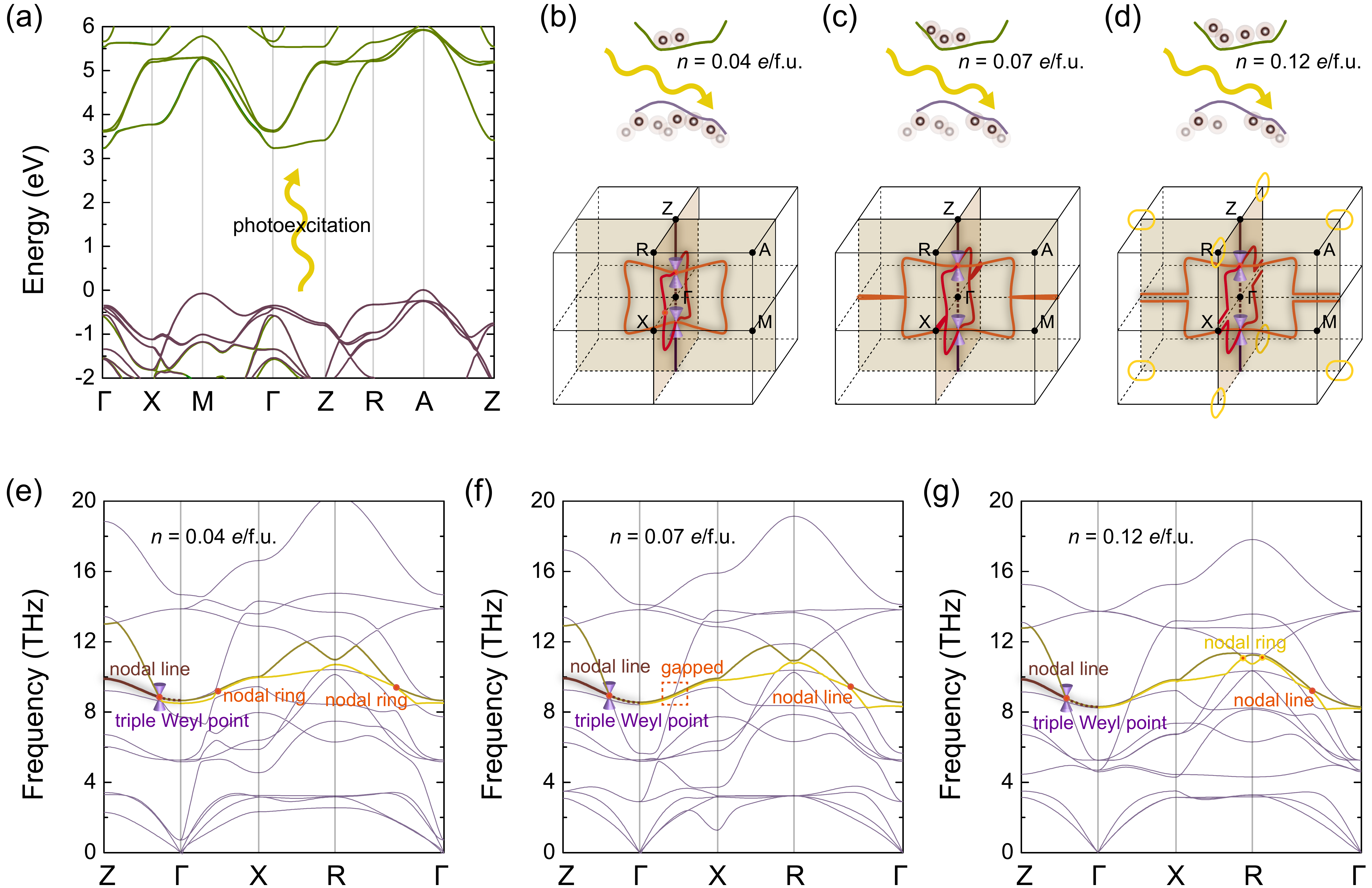}
\caption{\textbf{Nodal lines and nodal rings in tetragonal BaTiO$_3$ controlled by light.} (a) Schematic of the photoexcitation process in tetragonal BaTiO$_3$. (b)-(d) Nodal lines and nodal rings formed by the 10$^{th}$ and 11$^{th}$ phonon branches in the bulk Brillouin zone for photoexcited carrier concentrations of 0.04 $e$/f.u., 0.07 $e$/f.u., and 0.12 $e$/f.u. The corresponding phonon dispersions are shown in (e)-(g).}
\label{f3} 
\end{figure*}

As a large gap insulator, BaTiO$_3$ displays trivial topology in its electronic band structure [Fig.~\ref{f3}(a)]. However, there are three types of topological states in its phonon dispersion, which are nodal rings, nodal lines, and Weyl points. Since the unit cell of tetragonal BaTiO$_3$ has 5 atoms, there are 15 branches in the phonon spectrum. Band inversion between the 10$^{th}$ and 11$^{th}$ bands (labelled by increasing energy) along the X-$\Gamma$ high-symmetry line is protected by $M_{x}$ symmetry, which restricts the band crossing to a 1D continuous ring/line on the mirror-invariant plane \cite{Fang2016}. Therefore a nodal ring is formed on the $q_{x}=0$ plane. Because of the existence of an additional $C_{4z}$ symmetry, there is another nodal ring located on the $q_{y}=0$ plane, related to the one on the $q_{x}=0$ plane by $C_{4z}$ symmetry [orange circles in Fig.~\ref{f1}(e)]. The combination of $C_{4z}$ and mirror symmetries also brings out another type of topological phonon in BaTiO$_3$: an endless nodal line along the $\Gamma$-Z direction [brown line in Fig.~\ref{f1}(e)]. Intersection points between these two nodal rings and the nodal line form two triple Weyl points located along the Z'(0,0,-$\pi$)-$\Gamma$-Z(0,0,$\pi$) line. %In addition to the nodal rings, nodal line and triple Weyl phonons, a
Another topological feature of the band crossings between the 10$^{th}$ and 11$^{th}$ bands are eight Weyl points on the $q_{z}=\pi$ plane [purple and yellow cones in Fig.~\ref{f1}(e)]. 
These eight Weyl points, located at generic momenta on the $q_{z}=\pi$ plane, are not protected by any crystalline symmetry due to the breaking of the $M_{z}$ symmetry in the tetragonal lattice, so we need to apply more advanced symmetry-based indicator theory to diagnose the topology.

To gain a better understanding of the topological nature of the eight Weyl points present in photoexcited tetragonal BaTiO$_3$, we explore the existence of Weyl points from the perspective of topology as shown in Fig.~\ref{f2}(a), diagnosing topological degeneracies by using symmetry-based indicators from subgroups \cite{Po2017,Song2018b,Kruthoff2017}. After obtaining the symmetry data at $\Gamma$, M, and X, we note that the symmetry-indicator group for space group No.~99 is a trivial one, and the symmetry data at high-symmetry momenta does not satisfy the compatibility condition. Thus, we need to find a subgroup that has a nontrivial symmetry-based indicator group and satisfies the compatibility condition \cite{Zhang2020,Zhang2019c}. Space group No.~99 has five subgroups with the same size of unit cell, namely No.~75 ($P4$), No.~25 ($Pmm2$), No.~6 ($Pm$), No.~3 ($P2$), and No.~1 ($P1$). Among them, No.~3 is the maximal subgroup that has a nontrivial symmetry-based indicator group $\mathbb{Z}_{2}$ and satisfies the compatibility condition.
Thus, subgroup No.~3 is used for further diagnosis. Space group No.~3 only has a generator of $C_{2z}$ rotation symmetry, so the symmetry-based indicator group $\mathbb{Z}_{2}$ corresponds to the $z_{2}$ Berry phase of the loop enclosing half of the $q_{z}=0$ [X($\pi$,0,0)-M($\pi$,$\pi$,0)-M'(-$\pi$,$\pi$,0)-X'(-$\pi$,0,0)-X($\pi$,0,0)] or $q_{z}=\pi$ [R($\pi$,0,$\pi$)-A($\pi$,$\pi$,$\pi$)-A'(-$\pi$,$\pi$,$\pi$)-R'(-$\pi$,0,$\pi$)-R($\pi$,0,$\pi$)] plane. If $z_{2}=0$, there will be 0 $mod$ 4 Weyl points on the $q_{z}=0$ and $q_{z}=\pi$ planes. Our calculations deliver eight Weyl points in total, corresponding to the case $\mathbb{Z}_{2}=0$. Furthermore, considering the additional mirror symmetries in space group No.~99, the eight Weyl points on the $q_{z}=\pi$ plane are related by $M_{x}$ and $M_{y}$ symmetries: four of them with left-hand chirality, while the other four with right-hand chirality.

\subsection*{Controlling topological phonons by light}

% Additionally, changing the carrier concentration $n$ induced by photoexcitation leads to changes in the potential energy experienced by the ions, which in turn changes the interatomic force constants and the phonon dispersion. This implies that topological phonons can be controlled by light, as shown in Fig.~\ref{f3}(b)-(d). 

As discussed above, both photoexcitation with $0.04<n<0.10$\,$e$/f.u. and temperatures between $278$-$393$\,K can stabilize the tetragonal phase of BaTiO$_3$, as shown in Figs.~\ref{f1}(b) and (d). However, a main difference is that without photoexcitation, the long-range Coulomb interaction leads to a large LO-TO splitting \cite{Zhong1994a}, which tends to lift band degeneracies, as shown in Fig.~\ref{f1}(b). Consequently, the eight conventional Weyl points and the two triple Weyl points that we find in illuminated BaTiO$_3$ are not observed in temperature-stabilized tetragonal BaTiO$_3$ at 300 K (although the presence of nodal-line and nodal-ring phonons is confirmed in the Supplementary Materials). In contrast, under illumination, the photoexcited carriers lead to a strong free-carrier screening that suppresses the LO-TO energy splitting and facilitates the appearance of Weyl phonons in tetragonal BaTiO$_3$. Another advantage of the photoexcitation route is that, as discussed next, it can drive the phononic system between different topological quantum states, including switching between nodal-ring and nodal-line phonons, controlling the position of Weyl points in momentum space, and creating and annihilating these topological states.

%Under illumination, photoexcited electrons are occupying the bottom of the conduction band, creating equal number of holes on the top of the valence band. As shown in the photoexcitation scheme Fig.~\ref{f3}(a)-(d), by increasing photoexcited electron-hole pairs, more electrons are promoted to the conduction bands, and the interactions between electrons and ions are perturbed via electron density change, which indirectly changes the interatomic force constants. Therefore, photoexcitation can be used to manipulate the topological phonons. To demonstrate this explicitly, we calculate the phonon dispersions at different photoexcited carrier concentrations $n$.

As discussed above, changing the photoexcited carrier density brings about not only the transition between nodal rings and nodal lines, but also the creation of new nodal rings. As shown in Fig.~\ref{f3}(b)-(c), increasing the photoexcited carrier density from $n=0.04$ $e$/f.u. to $n=0.07$ $e$/f.u. drives the two nodal rings to become two nodal lines, and the corresponding phonon spectra are shown in Fig.~\ref{f3}(e)-(f). Further increasing $n$ to $0.12$ $e$/f.u. (under which conditions the tetragonal structure becomes cubic) leads to the formation of two nodal rings around the R point on the $q_{x}=0$ and $q_{y}=0$ planes, as shown in Fig.~\ref{f3}(d) and (g). Thus, by changing $n$, nodal rings in the phonon spectra can be created or transformed into nodal lines. 

\begin{figure*}[h]
\centering
\includegraphics[width=\linewidth]{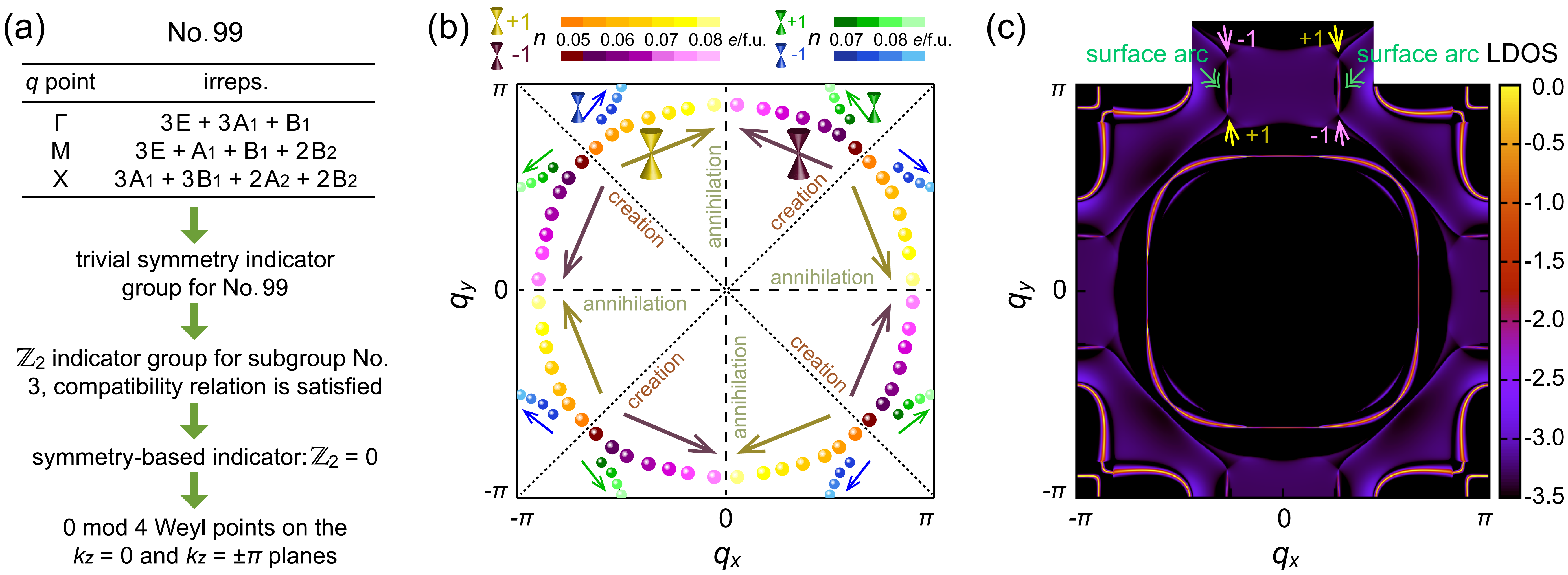}
\caption{\textbf{Weyl points in tetragonal BaTiO$_3$ controlled by light.} (a) Diagnosis process for the eight Weyl points in BaTiO$_3$ using symmetry-based indicator theory for the lowest 10 bands. (b) Evolution of the Weyl points on the $q_z$ = $\pi$ plane with increasing $n$ from $0.050$ to $0.085$ $e$/f.u. (c) Phonon surface arcs for a phonon frequency of 11.42 THz at $n=0.07$ $e$/f.u.} 
\label{f2} 
\end{figure*}

In addition to nodal lines and nodal rings, Weyl phonons are also sensitive to the photoexcited carrier concentration $n$. Fig.~\ref{f2}(b) shows the evolution of eight pairs of Weyl points on the $q_z=\pi$ plane when increasing $n$ from $0.050$ to $0.085$ $e$/f.u. (four extra pairs of Weyl points emerge when $0.0695<n<0.0825$ $e$/f.u.). Below $0.050$ $e$/f.u., BaTiO$_3$ exhibits no Weyl points between the 10$^{th}$ and 11$^{th}$ branches in its phonon spectrum. At about $n = 0.050$ $e$/f.u., four pairs of Weyl points are created around the $q_{x}=q_{y}$ and $q_{x}=-q_{y}$ planes [orange and maroon dots in Fig.~\ref{f2}(b)]. As $n$ increases, the Weyl points in each pair with different chirality move away from each other and head to the $q_{x}=0$ and $q_{y}=0$ planes [yellow and purple arrows in Fig.~\ref{f2}(b)]. 
Once $n$ reaches $0.0815$ $e$/f.u, Weyl points meet on the $q_{x}=0$ and $q_{y}=0$ planes and annihilate in pairs [light yellow and light magenta dots in Fig.~\ref{f2}(b)]. 
In addition, another four pairs of Weyl points of opposite chirality are created around the $q_{x}=q_{y}$ and $q_{x}=-q_{y}$ planes at $n=0.0695$ $e$/f.u., and annihilate in pairs on the $q_{x}=\pi$ and $q_{y}=\pi$ planes when $n$ reaches $0.0825$ $e$/f.u. [blue and green dots in Fig.~\ref{f2}(b)]. Since no band inversion happens at any high-symmetry points in this process, the Berry phase of the loop R-A-A'-R'-R on the $q_{z}=\pi$ plane remains zero, which is consistent with 0 $mod$ 4 Weyl points on the $q_{z}=\pi$ plane with $\mathbb{Z}_{2}=0$ [Fig.~\ref{f2}(a)].

To have a better understanding of the Weyl phonons in tetragonal BaTiO$_3$, we calculate the surface local density of states (LDOS) from the imaginary part of the surface Green's function \cite{Wu2018}. The surface LDOS in Fig.~\ref{f2}(c) is calculated along the (001) direction at $11.42$\,THz and $n=0.07$ $e$/f.u. The four pairs of Weyl points connect via surface arcs crossing the Brillouin zone boundaries, so that each surface arc starts from one Weyl point with a positive monopole charge and ends at another with a negative one. Overall, by modulating $n$ we can control the creation and annihilation of Weyl points, and the length of surface arcs in tetragonal BaTiO$_3$.

%Moreover, high temperature enhances phonon anharmonicity and electron-phonon coupling, and results in phonon linewidth broadening. {\color{red} Although for experimental detection of the phonon spectra, a broader phonon linewidth is easier to observe, it makes the phonon surface states merge in the bulk states (I will move surface states in SI)}. \textcolor{blue}{Comment: For experiments detecting the phonon spectra, it is better to have a broader bandwith.}

%Besides photoexcitation, light can provide an extra degree of tunability, which is more experimentally feasible compared to heating up, growth on substrate to create strain or doping. For instance, In addition, although increasing temperature to 300 K can stabilize the tetragonal phase, the long-range Coulomb interactions that lead to LO-TO splitting tend to lift the band degeneracy, as shown in Fig.~\ref{f1}(b), and no Weyl phonons are observed at 300 K. {\color{red} SI shows the nodal-line/ring phonons of RT BTO.} 

\subsection*{Topological phonons in PbTiO$_3$}

%SrTiO$_3$ is another well-studied perovskite that has a different phase diagram compared to BaTiO$_3$. At low temperatures it exhibits a centrosymmetric tetragonal phase of space group $I4/mcm$ (its phonon dispersion is shown in the Supplementary materials), and it undergoes a phase transition to a cubic phase above $106$\,K. 

Multiple types of topological phonons can also be found in other perovskites with the same $P4mm$ space group, and here we study PbTiO$_3$ and SrTiO$_3$ as additional examples.  %We find that upon illumination, the $P4mm$ phase can also be dynamically stabilized in SrTiO$_3$ when $n>0.075$ $e$/f.u. 

%The low-temperature phase of SrTiO$_3$ has a space group of $I4/mcm$ with octahedral distortions while retaining a centrosymmetric structure. At 106 K, it undergoes a phase transition to the cubic phase. In addition to temperature-induced cubic phase, the centrosymmetric structure can also undergo a ferroelectric phase transition to space group $P4mm$ under a small amount of negative strain.
%Recent experiments have shown that photoexcitation of lattice vibrations can induce polar order in SrTiO$_3$, which is metastable even for temperature above 290 K and can persist for hours after the pump \cite{Nova2019}. Moreover, second harmonic signal has indicated that the photo-induced ferroelectric phase transition originates from the hardening of low-frequency vibration modes \cite{Nova2019}.

\begin{figure*}[h]
\centering
\includegraphics[width=1\linewidth]{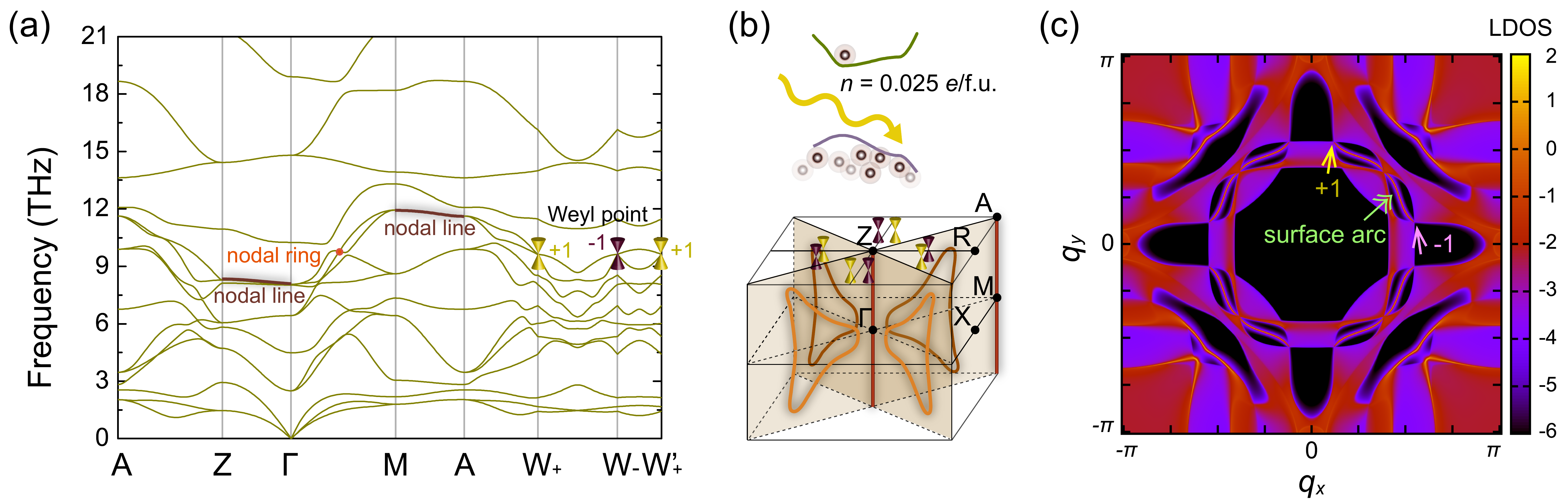}
\caption{\textbf{Topological phonons in tetragonal PbTiO$_3$.} (a) Phonon dispersion of tetragonal $P4mm$ PbTiO$_3$ at $n=0.025$ $e$/f.u. (b) Bulk Brillouin zone at $n=0.025$ $e$/f.u. with four nodal rings on the $q_{x}=q_{y}$ and $q_{x}=-q_{y}$ planes (orange circles), two nodal lines along the $\Gamma$-Z and M-A high-symmetry lines (brown line), and eight Weyl points located on $q_{z}=\pi$ plane (purple and yellow cones). All these three types of topological phonons are composed by the 10$^{th}$ and 11$^{th}$ bands. (c) Phonon surface arcs for a phonon frequency of 9.63 THz at $n=0.025$ $e$/f.u.}
\label{pto} 
\end{figure*}

PbTiO$_3$ has a tetragonal $P4mm$ structure for $T<600$\,K, and a cubic structure above that temperature. Athough the $P4mm$ phase has the lowest energy in a wide photoexcited density range $n<0.125$ $e$/f.u. \cite{Paillard2019}, it becomes dynamically unstable when $n>0.05$ $e$/f.u. (shown in the Supplementary Materials). We calculate the excited-state phonon dispersion at $n=0.025$ $e$/f.u., as shown in Fig.~\ref{pto}(a). Similar to BaTiO$_3$, PbTiO$_3$ also has nodal rings, nodal lines, and Weyl points between the 10$^{th}$ and 11$^{th}$ bands. But the triple Weyl points no longer exist as the nodal rings and the nodal line do not touch. The unique band crossings along the $\Gamma$-M high-symmetry line bring about four nodal rings on the $q_{x}=q_{y}$ and $q_{x}=-q_{y}$ planes protected by $M_{xy}$ and $M_{x\bar{y}}$ symmetries. The endless nodal lines along the $\Gamma$-Z and M-A directions are robust as well, as they are protected by the $C_{4z}$ symmetry.

The surface LDOS in Fig.~\ref{pto}(c) is calculated along the (001) direction at $9.63$\,THz with $n=0.025$ $e$/f.u. Four pairs of Weyl points, created at $n=0.005$ $e$/f.u., connect with each other via surface arcs crossing the $q_{x}=q_{y}$ and $q_{x}=-q_{y}$ planes. We note that the location of these four pairs of Weyl points can also be modulated by changing the photoexcitation density until the lattice becomes dynamically unstable for $n>0.05$ $e$/f.u.

\subsection*{Topological phonons in SrTiO$_3$}

\begin{figure*}[h]
\centering
\includegraphics[width=\linewidth]{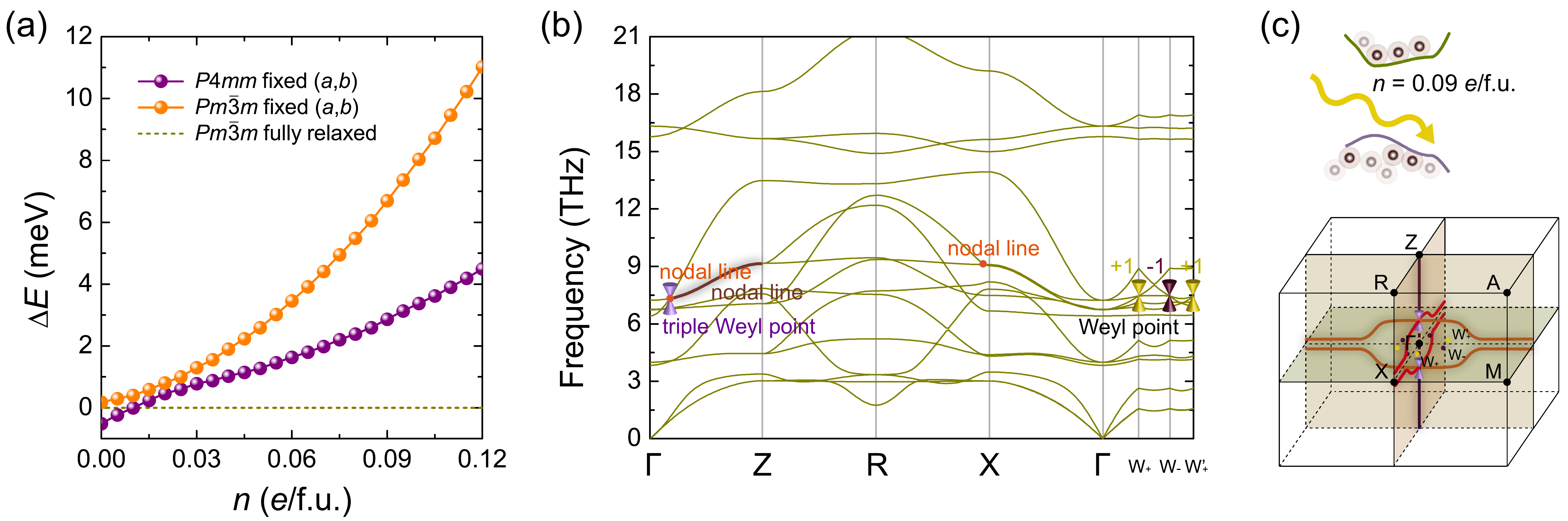}
\caption{\textbf{Topological phonons in tetragonal SrTiO$_3$.} (a) Relative energy difference between the $P4mm$ and $Pm\bar{3}m$ phases with fixed in-plane lattice constants $a$ and $b$, with the energy of the fully relaxed $Pm\bar{3}m$ phase set to be zero. (b) Phonon dispersion of tetragonal $P4mm$ SrTiO$_3$ at $n=0.09$ $e$/f.u. (c) Bulk Brillouin zone at $n=0.09$ $e$/f.u. with two nodal lines on the $q_{x}=0$ and $q_{y}=0$ planes (orange lines), one nodal line along the $\Gamma$-Z high-symmetry line (brown line), one pair of triple Weyl points at the intersection points between the three nodal lines (violet cones), and four pairs of Weyl points on the $q_z$ = $0$ plane (purple and yellow dots).
}
\label{sto} 
\end{figure*}

SrTiO$_3$ has a different phase diagram compared to BaTiO$_3$ and PbTiO$_3$. At low temperatures, it exhibits a centrosymmetric tetragonal phase of space group $I4/mcm$, and undergoes a phase transition to a cubic phase above $106$\,K. The $P4mm$ phase can only be accessed with small negative strains \cite{Dieguez2005,Ni2011}. The in-plane lattice constants $a$ and $b$ of $P4mm$ SrTiO$_3$ increase upon photoexcitation. Therefore by fixing $a$ and $b$ to the ones in the dark, we can induce in-plane negative strain in illuminated SrTiO$_3$. This can be realized experimentally by growing SrTiO$_3$ on an appropriate substrate. As shown in Fig.~\ref{sto}(a), after fixing the in-plane lattice constants $a$ and $b$ and relaxing only the out-of-plane lattice constant $c$ in the $P4mm$ phase, we find that the tetragonal phase can have a lower energy than the cubic $Pm\bar{3}m$ phase. 
%(technically the out-of-plane lattice constant $c$ of the $Pm\bar{3}m$ phase cannot be relaxed as it is a cubic phase).}

$P4mm$ SrTiO$_3$ exhibits imaginary phonon modes in the dark, while adding a photoexcited charge density of $n=0.09$ $e$/f.u. stabilizes the structure, as shown in Fig.~\ref{sto}(b). Similar to BaTiO$_3$, the $P4mm$ phase of SrTiO$_3$ has four nodal lines in total on the $q_{x}=0$ and $q_{y}=0$ mirror planes and one nodal line along the $\Gamma$-Z high-symmetry line in the bulk Brillouin zone. The intersection points of the three nodal lines form one pair of triple Weyl points. Different from BaTiO$_3$, the four pairs of Weyl points are located on the $q_{z}=0$ plane rather than the $q_{z}=\pi$ plane. Weyl phonons in SrTiO$_3$ can also be manipulated by photoexcitation, and the critical $n$ for the annihilation of the Weyl points on the $q_{x}=0$ and $q_{y}=0$ planes is 0.105 $e$/f.u., which is similar to that in BaTiO$_3$. %Different to BaTiO$_3$, SrTiO$_3$ exhibits no Weyl points in its phonon spectrum under illumination. 

%{\color{red} I run phonon calculations at every 0.005 $e$/f.u. and do not find Weyl phonons, although it has very similar phonon dispersion to BTO.}
%The topological properties of SrTiO$_3$ can be manipulated as well. Increasing $n$ to 0.135 $e$/f.u. enlarges the nodal ring to cross the A point [Fig.~\ref{f5}(b)], turning the nodal-ring phonons into nodal-line phonons, which is similar to BaTiO$_3$. HoweverDifferent from BaTiO$_3$, no Weyl phonons can be observed in a wide range of photoexcited carrer concentrations $n$ as the pseudo band crossings between the branches 14 and 15 are too close to the $\Gamma$ point. 

% The difference between $P4mm$ SrTiO$_3$ and BaTiO$_3$ is that the radius of the nodal ring in SrTiO$_3$ is much larger than BaTiO$_3$. Consequently, more clearly visible surface states can be expected in Fig.~\ref{f5}(e). A hallmark of nodal-ring phonons is the presence of a two-dimensional drumhead-like surface states connecting the projected nodal points, which are the projections of the bulk nodal line onto the (010) surface. These surface states can be detected by He4 scattering and HEELS {\color{red} references for the experiments}. Fig.~\ref{f5}(f) shows the isofrequency surface at 16.27 THz, corresponding to the thin green line in Fig.~\ref{f5}(e). The surface states are across the whole Brillouin zone, making SrTiO$_3$ a better candidate for measuring topological surface states than BaTiO$_3$ and PbTiO$_3$ whose nodal-ring/line surface states are mostly hidden in the bulk states.

\section*{Ubiquity of topological phonons in perovskite oxides}

The existence and coexistence of different types of topological phonons, together with their tunability by light, occur in a variety of other structural phases of perovskites beyond the tetragonal phases investigated above. Topological phonons in the orthorhombic $Amm2$ and the rhombohedral $R3m$ phases of BaTiO$_{3}$ are shown in the Supplementary Materials. %Here we only highlight the simplest $P4mm$ phase of three prototype perovskites.
Furthermore, topological phonons in oxide perovskites can also be manipulated by other tuning parameters beyond photoexcitation, such as strain and temperature (for details, see the Supplementary Materials). Putting these insights together suggests that topological phonons are ubiquitous in the family of perovskite oxides, which, given their widespread use and versatility, provide a promising platform for exploring topological phonon physics and its interplay with other phases.

 % However I find that a new tetragonal phase with polar displacement (\textit{i.e.}, the same $P4mm$ space group with BTO and PTO) can be stabilized by photoexcitation. 
% However, the surface states of the new nodal lines ({\color{red} in SI}) are not as clear as those of the nodal rings. 

%Based on the ubiquity of topological phonons in our calculations, we propose that the broad family of perovskite oxides can provide additional examples of topological phonons. Additionally, it has been shown that photoexcitation provides a viable route to stabilizing multiple perovskite phases, from EuTiO$_3$ to BiFeO$_3$ \cite{Paillard2019,Nova2019,Ahn2019,Porer2019,Juraschek2017}. Therefore, it would be interesting to search for additional perovskites exhibiting topological phonons {\color{red} and their interactions with other quasiparticles such as electrons and magnons}.

Given the proposed ubiquity of topological phonons in perovskites, it should be possible to directly identify them experimentally by looking for ``isolated'' band crossings, and the prototype materials detailed above are promising starting candidates. Topological bulk phonons can be measured by inelastic neutron scattering \cite{Zhang2019c,He2020,Choudhury2008} and inelastic X-ray scattering \cite{Miao2018}, while the accompanying topological surface states can be detected by high resolution electron energy loss spectroscopy \cite{Jia2017}. For light-induced topological phonons, although it may be difficult to illuminate the sample to maintain a constant photoexcited carrier density, transient photoexcitation could be used to observe the evolution of nodal rings, nodal lines and Weyl points as a function of the time-dependent photoexcited carrier density.

In terms of phenomena, topological phonons with non-zero Berry curvature may contribute to the recently observed phonon thermal Hall effect in SrTiO$_3$ \cite{Yang2020,Li2020}. In addition, novel phonon-phonon and electron-phonon scattering mechanisms could arise from topological phonons, which may provide new insights on the enhanced superconductivity of SrTiO$_3$ \cite{Lee2014a,Ahadi2019}. The emergence of Weyl phonons may be accompanied by other unique physical properties such as a pseudogauge field with a one-way propagating bulk mode \cite{Roy2018,Jia2019,Peri2019}, topological negative refraction \cite{He2018}, and nonlinear acoustic/optical responses \cite{Kim2019a,Sukhachov2020}, which can offer new routes for designing novel technologies like light-controlled neuromorphic computing in phononic systems \cite{Farmakidis2019,Feldmann2019}. Our work shows that oxide perovskites provide a promising platform to explore all of these phenomena and applications.

%branches 14 and 15 are completely gapped. The underlying physics is that the ferroelectric phase transition in PbTiO$_3$ involves much larger polar displacements of Ti and O atoms along the $z$ direction compared to BaTiO$_3$, which significantly breaks the mirror symmetry on the $q_z$ = 0 plane. Consisted of branches 13 and 14, four nodal-rings are formed on the plane XRAM. 
%\textcolor{blue}{How to describe operators: maybe we can use symbols like $C_{4z}$, $M_{y}$, $M_{xy}$, $M_{x,\bar{y}}$, or $C_{4}^{001}$, $M^{110}$, $M^{1\bar{1}0}$.} 
%Another four nodal rings are formed on the (100) and (010) mirror planes. The robust nodal line along the $\Gamma$-Z high-symmetry line is still there.

\section*{Conclusion}

We find that topological phonons are ubiquitous in oxide perovskites and that photoexcitation provides a promising route for their manipulation. As examples, the noncentrosymmetric tetragonal phases of three oxide perovskites (BaTiO$_3$, PbTiO$_3$, and SrTiO$_3$) exhibit topological nodal rings, nodal lines, and Weyl points in their phonon spectra. %Additionally, these topological phonons can be controlled by the photoexcited carrier density, which provides a new route to controlling topological phonon states. 
Remarkably, we find that photoexcitation is the only way to obtain Weyl phonons in tetragonal BaTiO$_{3}$, since the thermally-stabilized tetragonal phase has a large LO-TO energy splittting in the phonon spectrum that prevents band crossings. By contrast, photoexcited carriers screen long-range interactions, suppressing the LO-TO energy splitting and facilitating the band crossings. We also find that the photoexcited carrier density can be used to tune the creation and annihilation of Weyl points and nodal rings/lines without any associated structural phase transitions. Topological phonons in oxide perovskites provide a promising platform to study physical phenomena ranging from the phonon Hall effect to superconductivity, and may also offer new technological opportunities such as the realization of controllable topological quantum states for neuromorphic computing.

%Therefore studying topological phonons provides insights on understanding the physical process of multi-phase of perovskites including the evolution of Berry phase. 

% the pseudogauge fields can be generated in Weyl metamaterials: The chiral zeroth Landau level, which is a one-way propagative bulk mode,

%In illuminated perovskite systems, photoexcitation can induce eight phononic Weyl points and switch on/off nodal-ring/line phonons. The existence of Weyl phonons can be understood by symmetry-based indicator theory, while the nodal-line/ring phonons are protected by the (110) and (1$\bar{1}$0) mirror planes in perovskite systems. The light-triggered topological switch is a promising tool to control the exotic topological surface states. 

\section*{Methods}

Density functional theory (DFT) calculations are performed using the Vienna \textit{ab initio} simulation package ({\sc vasp}) with the projector-augmented-wave potential method \cite{Kresse1996,Kresse1996a}. We use the generalized gradient approximation (GGA) with the Perdew-Burke-Ernzerhof parameterization revised for solids (PBEsol) as the exchange-correlation functional \cite{Perdew2008}. A plane-wave basis set is employed with a kinetic energy cutoff of $800$\,eV and a $7\times7\times7$ \textbf{k}-mesh during structural relaxation, which is stopped when forces are below 10$^{-3}$ eV/\AA. The band structure of BaTiO$_3$ is calculated using the HSE06 hybrid functional in the presence of spin-orbit coupling \cite{HSE1}. Photoexcited carriers are simulated by promoting electrons from high-energy valence band states to low-energy conduction band states. This $\Delta$ self-consistent field ($\Delta$SCF) method introduces non-interacting electron-hole pairs by changing the occupation numbers of the Kohn-Sham orbitals \cite{Jones1989,Goerling1996,Hellman2004,Peng2020}, and is computationally less demanding compared to other approaches like constrained density functional theory \cite{Mauri1995} and excited-state force calculations \cite{Ismail-Beigi2003}. Nevertheless, it gives consistent phonon spectra compared with those obtained with constrained DFT \cite{Paillard2019}, as shown in the Supplementary Materials. The occupancies are fixed with a smearing of $0.01$\,eV.

The crystal structures in the dark and under photoexcitation are fully relaxed before the phonon calculations. We calculate the force constants with density functional perturbation theory (DFPT) \cite{DFPT} in a $2\times2\times2$ supercell with a $5\times5\times5$ \textbf{k}-mesh using {\sc vasp}. The phonon dispersion is then obtained using {\sc phonopy} \cite{Togo2015}. We perform convergence tests on the supercell size between $2\times2\times2$, $3\times3\times3$, and $4\times4\times2$, all confirming the existence of Weyl points in illuminated BaTiO$_3$. 
%Hereafter we use the force constants of the $2\times2\times2$ supercell to construct the phonon tight-binding Hamiltonian for simplicity.
We calculate the chirality of Weyl points by employing the Wilson-loop method to calculate the Wannier charge center flow, $i.e.$, the monopole charge of a Weyl point \cite{Soluyanov2011,Yu2011}. The phonon surface states are obtained using surface Green's functions as implemented in {\sc WannierTools} \cite{Wu2018}. The finite-temperature phonon frequencies including anharmonic contributions are calculated using a self-consistent \textit{ab initio} lattice dynamical method \cite{Souvatzis2008,Alfe2009,Cai2018} in a $2\times2\times2$ supercell. The self-consistent cycle is terminated after 240 iterations when the difference in free energy is less than 0.2 meV, and the space group symmetry is enforced on the resulting force constants. The Born effective charges are calculated using DFPT to obtain the LO-TO splitting in the phonon dispersion in the dark \cite{Gajdos2006}. Under illumination, the photoexcited electrons screen long-range interactions and no LO-TO splitting occurs. 

% Your references go at the end of the main text, and before the
% figures.  For this document we've used BibTeX, the .bib file
% scibib.bib, and the .bst file Science.bst.  The package scicite.sty
% was included to format the reference numbers according to *Science*
% style.

%BibTeX users: After compilation, comment out the following two lines and paste in
% the generated .bbl file. 

%\bibliography{new}

\begin{thebibliography}{10}

\bibitem{Heikkila2011}
T.~T. Heikkil\"a, N.~B. Kopnin, G.~E. Volovik, Flat bands in topological media,
  {\it JETP Letters\/} {\bf 94}, 233-- (2011).

\bibitem{Son2013}
D.~T. Son, B.~Z. Spivak, Chiral anomaly and classical negative
  magnetoresistance of weyl metals, {\it Phys. Rev. B\/} {\bf 88}, 104412
  (2013).

\bibitem{Xiong2015}
J.~Xiong, S.~K. Kushwaha, T.~Liang, J.~W. Krizan, M.~Hirschberger, W.~Wang,
  R.~J. Cava, N.~P. Ong, {Evidence for the chiral anomaly in the Dirac
  semimetal Na$_3$Bi}, {\it Science\/} {\bf 350}, 413 (2015).

\bibitem{Weng2015b}
H.~Weng, Y.~Liang, Q.~Xu, R.~Yu, Z.~Fang, X.~Dai, Y.~Kawazoe, Topological
  node-line semimetal in three-dimensional graphene networks, {\it Phys. Rev.
  B\/} {\bf 92}, 045108 (2015).

\bibitem{Bernevig2018}
A.~Bernevig, H.~Weng, Z.~Fang, X.~Dai, {Recent Progress in the Study of
  Topological Semimetals}, {\it J. Phys. Soc. Jpn.\/} {\bf 87}, 041001 (2018).

\bibitem{Fang2015}
C.~Fang, Y.~Chen, H.-Y. Kee, L.~Fu, Topological nodal line semimetals with and
  without spin-orbital coupling, {\it Phys. Rev. B\/} {\bf 92}, 081201 (2015).

\bibitem{Wang2012a}
Z.~Wang, Y.~Sun, X.-Q. Chen, C.~Franchini, G.~Xu, H.~Weng, X.~Dai, Z.~Fang,
  {Dirac semimetal and topological phase transitions in ${A}_{3}$Bi
  ($A=\text{Na}$, K, Rb)}, {\it Phys. Rev. B\/} {\bf 85}, 195320 (2012).

\bibitem{Armitage2018}
N.~P. Armitage, E.~J. Mele, A.~Vishwanath, {Weyl and Dirac semimetals in
  three-dimensional solids}, {\it Rev. Mod. Phys.\/} {\bf 90}, 015001 (2018).

\bibitem{Xu2015}
S.-Y. Xu, I.~Belopolski, N.~Alidoust, M.~Neupane, G.~Bian, C.~Zhang, R.~Sankar,
  G.~Chang, Z.~Yuan, C.-C. Lee, S.-M. Huang, H.~Zheng, J.~Ma, D.~S. Sanchez,
  B.~Wang, A.~Bansil, F.~Chou, P.~P. Shibayev, H.~Lin, S.~Jia, M.~Z. Hasan,
  Discovery of a weyl fermion semimetal and topological fermi arcs, {\it
  Science\/} {\bf 349}, 613--617 (2015).

\bibitem{Weng2015}
H.~Weng, C.~Fang, Z.~Fang, B.~A. Bernevig, X.~Dai, Weyl semimetal phase in
  noncentrosymmetric transition-metal monophosphides, {\it Phys. Rev. X\/} {\bf
  5}, 011029 (2015).

\bibitem{Yang2015}
L.~X. Yang, Z.~K. Liu, Y.~Sun, H.~Peng, H.~F. Yang, T.~Zhang, B.~Zhou,
  Y.~Zhang, Y.~F. Guo, M.~Rahn, D.~Prabhakaran, Z.~Hussain, S.-K. Mo,
  C.~Felser, B.~Yan, Y.~L. Chen, Weyl semimetal phase in the
  non-centrosymmetric compound taas, {\it Nat Phys\/} {\bf 11}, 728--732
  (2015).

\bibitem{Yang2018}
S.-Y. Yang, H.~Yang, E.~Derunova, S.~S.~P. Parkin, B.~Yan, M.~N. Ali, Symmetry
  demanded topological nodal-line materials, {\it Advances in Physics: X\/}
  {\bf 3}, 1414631-- (2018).

\bibitem{He2017}
J.~He, X.~Li, P.~Lyu, P.~Nachtigall, Near-room-temperature chern insulator and
  dirac spin-gapless semiconductor: nickel chloride monolayer, {\it
  Nanoscale\/} {\bf 9}, 2246--2252 (2017).

\bibitem{Wu2017a}
F.~Wu, T.~Lovorn, A.~H. MacDonald, Topological exciton bands in moir\'e
  heterojunctions, {\it Phys. Rev. Lett.\/} {\bf 118}, 147401 (2017).

\bibitem{Li2016d}
F.-Y. Li, Y.-D. Li, Y.~B. Kim, L.~Balents, Y.~Yu, G.~Chen, Weyl magnons in
  breathing pyrochlore antiferromagnets, {\it Nature Communications\/} {\bf 7},
  12691-- (2016).

\bibitem{Stenull2016}
O.~Stenull, C.~L. Kane, T.~C. Lubensky, Topological phonons and weyl lines in
  three dimensions, {\it Phys. Rev. Lett.\/} {\bf 117}, 068001 (2016).

\bibitem{Liu2017a}
Y.~Liu, Y.~Xu, S.-C. Zhang, W.~Duan, Model for topological phononics and phonon
  diode, {\it Phys. Rev. B\/} {\bf 96}, 064106 (2017).

\bibitem{He2018}
H.~He, C.~Qiu, L.~Ye, X.~Cai, X.~Fan, M.~Ke, F.~Zhang, Z.~Liu, Topological
  negative refraction of surface acoustic waves in a weyl phononic crystal,
  {\it Nature\/} {\bf 560}, 61--64 (2018).

\bibitem{Zhang2018a}
T.~Zhang, Z.~Song, A.~Alexandradinata, H.~Weng, C.~Fang, L.~Lu, Z.~Fang,
  Double-weyl phonons in transition-metal monosilicides, {\it Phys. Rev.
  Lett.\/} {\bf 120}, 016401 (2018).

\bibitem{Miao2018}
H.~Miao, T.~T. Zhang, L.~Wang, D.~Meyers, A.~H. Said, Y.~L. Wang, Y.~G. Shi,
  H.~M. Weng, Z.~Fang, M.~P.~M. Dean, Observation of double weyl phonons in
  parity-breaking fesi, {\it Phys. Rev. Lett.\/} {\bf 121}, 035302 (2018).

\bibitem{Li2018a}
J.~Li, Q.~Xie, S.~Ullah, R.~Li, H.~Ma, D.~Li, Y.~Li, X.-Q. Chen, Coexistent
  three-component and two-component weyl phonons in tis, zrse, and hfte, {\it
  Phys. Rev. B\/} {\bf 97}, 054305 (2018).

\bibitem{Xia2019}
B.~W. Xia, R.~Wang, Z.~J. Chen, Y.~J. Zhao, H.~Xu, Symmetry-protected ideal
  type-ii weyl phonons in cdte, {\it Phys. Rev. Lett.\/} {\bf 123}, 065501
  (2019).

\bibitem{Zhang2019c}
T.~T. Zhang, H.~Miao, Q.~Wang, J.~Q. Lin, Y.~Cao, G.~Fabbris, A.~H. Said,
  X.~Liu, H.~C. Lei, Z.~Fang, H.~M. Weng, M.~P.~M. Dean, Phononic helical nodal
  lines with $\mathcal{PT}$ protection in ${\mathrm{mob}}_{2}$, {\it Phys. Rev.
  Lett.\/} {\bf 123}, 245302 (2019).

\bibitem{Liu2020}
Y.~Liu, X.~Chen, Y.~Xu, Topological phononics: From fundamental models to real
  materials, {\it Adv. Funct. Mater.\/} {\bf 30}, 1904784-- (2020).

\bibitem{Drozdov2015}
A.~P. Drozdov, M.~I. Eremets, I.~A. Troyan, V.~Ksenofontov, S.~I. Shylin,
  Conventional superconductivity at 203 kelvin at high pressures in the sulfur
  hydride system, {\it Nature\/} {\bf 525}, 73--76 (2015).

\bibitem{Ahadi2019}
K.~Ahadi, L.~Galletti, Y.~Li, S.~Salmani-Rezaie, W.~Wu, S.~Stemmer, Enhancing
  superconductivity in srtio\&lt;sub\&gt;3\&lt;/sub\&gt; films with strain,
  {\it Sci Adv\/} {\bf 5}, eaaw0120-- (2019).

\bibitem{Murakami2016}
S.~Murakami, A.~Okamoto, Thermal hall effect of magnons, {\it J. Phys. Soc.
  Jpn.\/} {\bf 86}, 011010-- (2016).

\bibitem{Chen2020}
J.-Y. Chen, S.~A. Kivelson, X.-Q. Sun, Enhanced thermal hall effect in nearly
  ferroelectric insulators, {\it Phys. Rev. Lett.\/} {\bf 124}, 167601 (2020).

\bibitem{Hamada2020}
M.~Hamada, S.~Murakami, Phonon rotoelectric effect, {\it Phys. Rev. B\/} {\bf
  101}, 144306 (2020).

\bibitem{Li2020}
X.~Li, B.~Fauqu\'e, Z.~Zhu, K.~Behnia, Phonon thermal hall effect in strontium
  titanate, {\it Phys. Rev. Lett.\/} {\bf 124}, 105901 (2020).

\bibitem{Cohen1992}
R.~E. Cohen, Origin of ferroelectricity in perovskite oxides, {\it Nature\/}
  {\bf 358}, 136--138 (1992).

\bibitem{Moritomo1997}
Y.~Moritomo, H.~Kuwahara, Y.~Tomioka, Y.~Tokura, Pressure effects on
  charge-ordering transitions in perovskite manganites, {\it Phys. Rev. B\/}
  {\bf 55}, 7549--7556 (1997).

\bibitem{Nahas2015}
Y.~Nahas, S.~Prokhorenko, L.~Louis, Z.~Gui, I.~Kornev, L.~Bellaiche, Discovery
  of stable skyrmionic state in ferroelectric nanocomposites, {\it Nature
  Communications\/} {\bf 6}, 8542-- (2015).

\bibitem{Das2019}
S.~Das, Y.~L. Tang, Z.~Hong, M.~A.~P. Goncalves, M.~R. McCarter, C.~Klewe,
  K.~X. Nguyen, F.~G\'omez-Ortiz, P.~Shafer, E.~Arenholz, V.~A. Stoica, S.-L.
  Hsu, B.~Wang, C.~Ophus, J.~F. Liu, C.~T. Nelson, S.~Saremi, B.~Prasad, A.~B.
  Mei, D.~G. Schlom, J.~\'Iniguez, P.~Garc\'ia-Fern\'andez, D.~A. Muller, L.~Q.
  Chen, J.~Junquera, L.~W. Martin, R.~Ramesh, Observation of room-temperature
  polar skyrmions, {\it Nature\/} {\bf 568}, 368--372 (2019).

\bibitem{Mahmoud2014}
A.~Mahmoud, A.~Erba, K.~E. El-Kelany, M.~R\'erat, R.~Orlando, Low-temperature
  phase of batio${}_{3}$: Piezoelectric, dielectric, elastic, and photoelastic
  properties from ab initio simulations, {\it Phys. Rev. B\/} {\bf 89}, 045103
  (2014).

\bibitem{Ahn2004}
K.~H. Ahn, T.~Lookman, A.~R. Bishop, Strain-induced metal-insulator phase
  coexistence in perovskite manganites, {\it Nature\/} {\bf 428}, 401--404
  (2004).

\bibitem{Shirako2009}
Y.~Shirako, H.~Kojitani, M.~Akaogi, K.~Yamaura, E.~Takayama-Muromachi,
  High-pressure phase transitions of carho3 perovskite, {\it Physics and
  Chemistry of Minerals\/} {\bf 36}, 455-- (2009).

\bibitem{Rusevich2019}
L.~L. Rusevich, G.~Zvejnieks, E.~A. Kotomin, M.~M. Krzmanc, A.~Meden, S.~Kunej,
  I.~D. Vlaicu, Theoretical and experimental study of (ba,sr)tio3 perovskite
  solid solutions and batio3/srtio3 heterostructures, {\it J. Phys. Chem. C\/}
  {\bf 123}, 2031--2036 (2019).

\bibitem{Paillard2019}
C.~Paillard, E.~Torun, L.~Wirtz, J.~\'I\~niguez, L.~Bellaiche, Photoinduced
  phase transitions in ferroelectrics, {\it Phys. Rev. Lett.\/} {\bf 123},
  087601 (2019).

\bibitem{Nova2019}
T.~F. Nova, A.~S. Disa, M.~Fechner, A.~Cavalleri, {{Metastable ferroelectricity
  in optically strained SrTiO$_3$}}, {\it Science\/} {\bf 364}, 1075-- (2019).

\bibitem{Ahn2019}
Y.~Ahn, A.~Pateras, S.~D. Marks, H.~Xu, T.~Zhou, Z.~Luo, Z.~Chen, L.~Chen,
  X.~Zhang, A.~D. DiChiara, H.~Wen, P.~G. Evans, Nanosecond optically induced
  phase transformation in compressively strained ${\mathrm{bifeo}}_{3}$ on
  ${\mathrm{laalo}}_{3}$, {\it Phys. Rev. Lett.\/} {\bf 123}, 045703 (2019).

\bibitem{Porer2019}
M.~Porer, M.~Fechner, M.~Kubli, M.~J. Neugebauer, S.~Parchenko, V.~Esposito,
  A.~Narayan, N.~A. Spaldin, R.~Huber, M.~Radovic, E.~M. Bothschafter, J.~M.
  Glownia, T.~Sato, S.~Song, S.~L. Johnson, U.~Staub, Ultrafast transient
  increase of oxygen octahedral rotations in a perovskite, {\it Phys. Rev.
  Research\/} {\bf 1}, 012005 (2019).

\bibitem{Juraschek2017}
D.~M. Juraschek, M.~Fechner, N.~A. Spaldin, Ultrafast structure switching
  through nonlinear phononics, {\it Phys. Rev. Lett.\/} {\bf 118}, 054101
  (2017).

\bibitem{KAY1947}
H.~F. KAY, R.~G. RHODES, Barium titanate crystals, {\it Nature\/} {\bf 160},
  126--127 (1947).

\bibitem{Souvatzis2008}
P.~Souvatzis, O.~Eriksson, M.~I. Katsnelson, S.~P. Rudin, Entropy driven
  stabilization of energetically unstable crystal structures explained from
  first principles theory, {\it Phys. Rev. Lett.\/} {\bf 100}, 095901 (2008).

\bibitem{Fang2016}
C.~Fang, H.~Weng, X.~Dai, Z.~Fang, Topological nodal line semimetals, {\it
  Chinese Physics B\/} {\bf 25}, 117106-- (2016).

\bibitem{Po2017}
H.~C. Po, A.~Vishwanath, H.~Watanabe, Symmetry-based indicators of band
  topology in the 230 space groups, {\it Nature Communications\/} {\bf 8}, 50--
  (2017).

\bibitem{Song2018b}
Z.~Song, T.~Zhang, C.~Fang, Diagnosis for nonmagnetic topological semimetals in
  the absence of spin-orbital coupling, {\it Phys. Rev. X\/} {\bf 8}, 031069
  (2018).

\bibitem{Kruthoff2017}
J.~Kruthoff, J.~de~Boer, J.~van Wezel, C.~L. Kane, R.-J. Slager, Topological
  classification of crystalline insulators through band structure
  combinatorics, {\it Phys. Rev. X\/} {\bf 7}, 041069 (2017).

\bibitem{Zhang2020}
T.~Zhang, L.~Lu, S.~Murakami, Z.~Fang, H.~Weng, C.~Fang, Diagnosis scheme for
  topological degeneracies crossing high-symmetry lines, {\it Phys. Rev.
  Research\/} {\bf 2}, 022066 (2020).

\bibitem{Zhong1994a}
W.~Zhong, R.~D. King-Smith, D.~Vanderbilt, Giant lo-to splittings in perovskite
  ferroelectrics, {\it Phys. Rev. Lett.\/} {\bf 72}, 3618--3621 (1994).

\bibitem{Wu2018}
Q.~Wu, S.~Zhang, H.-F. Song, M.~Troyer, A.~A. Soluyanov, {WannierTools: An
  open-source software package for novel topological materials}, {\it Computer
  Physics Communications\/} {\bf 224}, 405--416 (2018).

\bibitem{Dieguez2005}
O.~Di\'eguez, K.~M. Rabe, D.~Vanderbilt, First-principles study of epitaxial
  strain in perovskites, {\it Phys. Rev. B\/} {\bf 72}, 144101 (2005).

\bibitem{Ni2011}
L.~Ni, Y.~Liu, C.~Song, W.~Wang, G.~Han, Y.~Ge, First-principle study of
  strain-driven phase transition in incipient ferroelectric srtio3, {\it
  Physica B: Condensed Matter\/} {\bf 406}, 4145--4149 (2011).

\bibitem{He2020}
X.~He, D.~Bansal, B.~Winn, S.~Chi, L.~Boatner, O.~Delaire, Anharmonic
  eigenvectors and acoustic phonon disappearance in quantum paraelectric
  ${\mathrm{srtio}}_{3}$, {\it Phys. Rev. Lett.\/} {\bf 124}, 145901 (2020).

\bibitem{Choudhury2008}
N.~Choudhury, E.~J. Walter, A.~I. Kolesnikov, C.-K. Loong, Large phonon band
  gap in $\mathrm{Sr}\mathrm{Ti}{\mathrm{o}}_{3}$ and the vibrational
  signatures of ferroelectricity in $a\mathrm{Ti}{\mathrm{o}}_{3}$ perovskites:
  First-principles lattice dynamics and inelastic neutron scattering, {\it
  Phys. Rev. B\/} {\bf 77}, 134111 (2008).

\bibitem{Jia2017}
X.~Jia, S.~Zhang, R.~Sankar, F.-C. Chou, W.~Wang, K.~Kempa, E.~W. Plummer,
  J.~Zhang, X.~Zhu, J.~Guo, Anomalous acoustic plasmon mode from topologically
  protected states, {\it Phys. Rev. Lett.\/} {\bf 119}, 136805 (2017).

\bibitem{Yang2020}
Y.-f. Yang, G.-M. Zhang, F.-C. Zhang, Universal behavior of the thermal hall
  conductivity, {\it Phys. Rev. Lett.\/} {\bf 124}, 186602 (2020).

\bibitem{Lee2014a}
J.~J. Lee, F.~T. Schmitt, R.~G. Moore, S.~Johnston, Y.-T. Cui, W.~Li, M.~Yi,
  Z.~K. Liu, M.~Hashimoto, Y.~Zhang, D.~H. Lu, T.~P. Devereaux, D.-H. Lee,
  Z.-X. Shen, Interfacial mode coupling as the origin of the enhancement of tc
  in fese films on srtio3, {\it Nature\/} {\bf 515}, 245--248 (2014).

\bibitem{Roy2018}
S.~Roy, M.~Kolodrubetz, N.~Goldman, A.~G. Grushin, Tunable axial gauge fields
  in engineered weyl semimetals: semiclassical analysis and optical lattice
  implementations, {\it 2D Materials\/} {\bf 5}, 024001-- (2018).

\bibitem{Jia2019}
H.~Jia, R.~Zhang, W.~Gao, Q.~Guo, B.~Yang, J.~Hu, Y.~Bi, Y.~Xiang, C.~Liu,
  S.~Zhang, Observation of chiral zero mode in inhomogeneous three-dimensional
  weyl metamaterials, {\it Science\/} {\bf 363}, 148-- (2019).

\bibitem{Peri2019}
V.~Peri, M.~Serra-Garcia, R.~Ilan, S.~D. Huber, Axial-field-induced chiral
  channels in an acoustic weyl system, {\it Nature Physics\/} {\bf 15},
  357--361 (2019).

\bibitem{Kim2019a}
J.~Kim, K.-W. Kim, D.~Shin, S.-H. Lee, J.~Sinova, N.~Park, H.~Jin, Prediction
  of ferroelectricity-driven berry curvature enabling charge- and
  spin-controllable photocurrent in tin telluride monolayers, {\it Nature
  Communications\/} {\bf 10}, 3965-- (2019).

\bibitem{Sukhachov2020}
P.~O. Sukhachov, H.~Rostami, Acoustogalvanic effect in dirac and weyl
  semimetals, {\it Phys. Rev. Lett.\/} {\bf 124}, 126602 (2020).

\bibitem{Farmakidis2019}
N.~Farmakidis, N.~Youngblood, X.~Li, J.~Tan, J.~L. Swett, Z.~Cheng, C.~D.
  Wright, W.~H.~P. Pernice, H.~Bhaskaran, Plasmonic nanogap enhanced
  phase-change devices with dual electrical-optical functionality, {\it Sci
  Adv\/} {\bf 5}, eaaw2687-- (2019).

\bibitem{Feldmann2019}
J.~Feldmann, N.~Youngblood, C.~D. Wright, H.~Bhaskaran, W.~H.~P. Pernice,
  All-optical spiking neurosynaptic networks with self-learning capabilities,
  {\it Nature\/} {\bf 569}, 208--214 (2019).

\bibitem{Kresse1996}
G.~Kresse, J.~Furthm\"uller, {Efficient iterative schemes for \textit{ab
  initio} total-energy calculations using a plane-wave basis set}, {\it Phys.
  Rev. B\/} {\bf 54}, 11169--11186 (1996).

\bibitem{Kresse1996a}
G.~Kresse, J.~Furthm\"uller, Efficiency of ab-initio total energy calculations
  for metals and semiconductors using a plane-wave basis set, {\it
  Computational Materials Science\/} {\bf 6}, 15 - 50 (1996).

\bibitem{Perdew2008}
J.~P. Perdew, A.~Ruzsinszky, G.~I. Csonka, O.~A. Vydrov, G.~E. Scuseria, L.~A.
  Constantin, X.~Zhou, K.~Burke, {{Restoring the Density-Gradient Expansion for
  Exchange in Solids and Surfaces}}, {\it Phys. Rev. Lett.\/} {\bf 100}, 136406
  (2008).

\bibitem{HSE1}
J.~Heyd, G.~E. Scuseria, M.~Ernzerhof, {Hybrid functionals based on a screened
  Coulomb potential}, {\it J. Chem. Phys.\/} {\bf 118}, 8207 (2003).

\bibitem{Jones1989}
R.~O. Jones, O.~Gunnarsson, The density functional formalism, its applications
  and prospects, {\it Rev. Mod. Phys.\/} {\bf 61}, 689--746 (1989).

\bibitem{Goerling1996}
A.~G\"orling, Density-functional theory for excited states, {\it Phys. Rev.
  A\/} {\bf 54}, 3912--3915 (1996).

\bibitem{Hellman2004}
A.~Hellman, B.~Razaznejad, B.~I. Lundqvist, Potential-energy surfaces for
  excited states in extended systems, {\it J. Chem. Phys.\/} {\bf 120},
  4593--4602 (2004).

\bibitem{Peng2020}
B.~Peng, H.~Zhang, W.~Chen, B.~Hou, Z.-J. Qiu, H.~Shao, H.~Zhu, B.~Monserrat,
  D.~Fu, H.~Weng, C.~M. Soukoulis, Sub-picosecond photo-induced displacive
  phase transition in two-dimensional mote2, {\it npj 2D Materials and
  Applications\/} {\bf 4}, 14-- (2020).

\bibitem{Mauri1995}
F.~Mauri, R.~Car, First-principles study of excitonic self-trapping in diamond,
  {\it Phys. Rev. Lett.\/} {\bf 75}, 3166--3169 (1995).

\bibitem{Ismail-Beigi2003}
S.~Ismail-Beigi, S.~G. Louie, Excited-state forces within a first-principles
  green's function formalism, {\it Phys. Rev. Lett.\/} {\bf 90}, 076401 (2003).

\bibitem{DFPT}
S.~Baroni, S.~de~Gironcoli, A.~Dal~Corso, P.~Giannozzi, Phonons and related
  crystal properties from density-functional perturbation theory, {\it Rev.
  Mod. Phys.\/} {\bf 73}, 515-562 (2001).

\bibitem{Togo2015}
A.~Togo, I.~Tanaka, {First principles phonon calculations in materials
  science}, {\it Scripta Materialia\/} {\bf 108}, 1-5 (2015).

\bibitem{Soluyanov2011}
A.~A. Soluyanov, D.~Vanderbilt, Computing topological invariants without
  inversion symmetry, {\it Phys. Rev. B\/} {\bf 83}, 235401 (2011).

\bibitem{Yu2011}
R.~Yu, X.~L. Qi, A.~Bernevig, Z.~Fang, X.~Dai, Equivalent expression of
  ${\mathbb{z}}_{2}$ topological invariant for band insulators using the
  non-abelian berry connection, {\it Phys. Rev. B\/} {\bf 84}, 075119 (2011).

\bibitem{Alfe2009}
D.~Alf\`e, Phon: A program to calculate phonons using the small displacement
  method, {\it Computer Physics Communications\/} {\bf 180}, 2622--2633 (2009).

\bibitem{Cai2018}
B.~Cai, X.~Chen, M.~Xie, S.~Zhang, X.~Liu, J.~Yang, W.~Zhou, S.~Guo, H.~Zeng, A
  class of pb-free double perovskite halide semiconductors with intrinsic
  ferromagnetism, large spin splitting and high curie temperature, {\it Mater.
  Horiz.\/} {\bf 5}, 961--968 (2018).

\bibitem{Gajdos2006}
M.~Gajdo\v{s}, K.~Hummer, G.~Kresse, J.~Furthm\"uller, F.~Bechstedt, Linear
  optical properties in the projector-augmented wave methodology, {\it Phys.
  Rev. B\/} {\bf 73}, 045112 (2006).

\end{thebibliography}

\bibliographystyle{Science}

\section*{Acknowledgments}
% Include acknowledgments of funding, any patents pending, where raw data for the paper are deposited, etc.

The authors gratefully acknowledge helpful discussions with Danny Bennett (University of Cambridge) on the phase diagram of oxide perovskites, Prof. Guntars Zvejnieks (University of Latvia) on the crystal structures of BaTiO$_3$, Bowen Hou and Prof. Hao Zhang (Fudan University) on the $\Delta$SCF method, Dr Bo Cai (Nanjing University of Science and Technology) on finite-temperature lattice dynamics, and Prof. Stefano Baroni (SISSA) on the screening effect of photoexcited carriers. B.P. and B.M. acknowledge support from the Winton Programme for the Physics of Sustainability, and B.M. also acknowledges support from the Gianna Angelopoulos Programme for Science, Technology, and Innovation. Y.H. acknowledges support from EPSRC grant EP/R512461/1 and Trinity College Henry-Barlow Scholarship. S.M. and T.T.Z. acknowledge support from the Tokodai Institute for Element Strategy (TIES) funded by MEXT Elements Strategy Initiative to Form Core Research Center, and S.M. also acknowledges support by JSPS KAKENHI Grant Number JP18H03678. The calculations were performed using resources provided by the Cambridge Tier-2 system operated by the University of Cambridge Research Computing Service (http://www.hpc.cam.ac.uk) and funded by EPSRC Tier-2 capital grant EP/P020259/1, and also with computational support from the UK Materials and Molecular Modelling Hub, which is partially funded by EPSRC (EP/P020194), for which access was obtained via the UKCP consortium and funded by EPSRC grant ref EP/P022561/1.

\noindent \textbf{Competing Interests:} The authors declare that they have no competing interests.

\noindent \textbf{Author Contributions:} B.M., T.T.Z. and B.P. devised the project idea. B.P., Y.H. and T.T.Z. performed the calculations. B.P, T.T.Z. and B.M. prepared the main part of the manuscript. B.P., Y.H., S.M., T.T.Z. and B.M. discussed the results and the ideas for their analysis and edited the manuscript.

\noindent \textbf{Data Availability:} All data needed to evaluate the conclusions in the paper are present in the paper and/or the Supplementary Materials. Additional data related to this paper may be requested from the authors.

%Here you should list the contents of your Supplementary Materials -- below is an example. 
%You should include a list of Supplementary figures, Tables, and any references that appear only in the SM. 
%Note that the reference numbering continues from the main text to the SM.
% In the example below, Refs. 4-10 were cited only in the SM.     
\section*{Supplementary materials}
%Materials and Methods\\
Supplementary Text\\
Figs. S1 to S9\\
%Tables S1 to S4\\
% References \textit{(4-10)}

% For your review copy (i.e., the file you initially send in for
% evaluation), you can use the {figure} environment and the
% \includegraphics command to stream your figures into the text, placing
% all figures at the end.  For the final, revised manuscript for
% acceptance and production, however, PostScript or other graphics
% should not be streamed into your compliled file.  Instead, set
% captions as simple paragraphs (with a \noindent tag), setting them
% off from the rest of the text with a \clearpage as shown  below, and
% submit figures as separate files according to the Art Department's
% instructions.

\end{document}